\documentclass[doublecol,a4paper]{epl2}
\usepackage{amsmath}
\usepackage{amsfonts}
\usepackage{amssymb}

\title{Dynamics of a colloidal glass during stress-mediated structural arrest}
\shorttitle{Stress activated dynamics during structural arrest} 

\author{A. Negi \inst{1} \and C. Osuji\inst{1}}

\institute{
  \inst{1} Department of Chemical Engineering, Yale University, New Haven CT 06511\\
}

\abstract{We employ parallel superposition rheology to study the
dynamics of an aging colloidal glass in the presence of a mean field
stress $\sigma_m$. Over a range of intermediate stresses, the loss
modulus exceeds the storage modulus at short times but develops a
maximum concomitant with a crossover between the two as the system
ages. This is attended by a narrowing of the loss peak on increasing
stress. We show that this feature is characteristic of the
structural arrest in these materials, which is made observable on
reasonable timescales by the activating influence of the stress. The
arrest time displays an exponential dependence on inverse stress.
These results provide experimental validation of the role of stress
as an effective temperature in soft glassy systems as has been
advanced in recent theoretical frameworks. }

\begin{document}

\maketitle The thermalizing effects of shear have been
well recognized in colloidal systems
\cite{Thirumalai_PRE1997,Wyss_SRFS2007,Liu_PRL2007,Ilg2007} where an
applied shear stress lowers energetic barriers to particle motion,
or an imposed shear rate provides higher particle mobility than
expected based on the system (bath) temperature. In such scenarios,
accelerated dynamics are found in equilibrium systems under shear
such as non-Brownian particulate suspensions \cite{Pine_Nature2005}
and polymers in non-linear creep flow where time-stress
superposition is observed \cite{Schapery_time_stress1969}. However
the effects of imposed stress on the dynamics of out of equilibrium
glassy materials, that is, in systems which display aging, has not
been explored experimentally in great detail. In the presence of an
applied stress, such systems, for example gels, pastes, emulsions
and slurries, are known to display a viscosity bifurcation
\cite{Coussot_PRE2002,Coussot_Pastes2007}. Application of a
sub-critical stresses $\sigma<\sigma_c$ results in an eventual
immobile state as the shear rate developed in response to the stress
decreases due to an ever increasing viscosity as the system ages or
structures over time. By contrast, application of a larger than
critical stress produces a quasi-steady state response with a well
defined shear rate and finite viscosity. These respective displays
are due to the dominance of either aging or rejuvenation in the
presence of the applied stresses, where rejuvenation in this sense
should be understood simply as a decrease of the characteristic
relaxation time of the system, rather than a strict path reversal of
aging.

Here, we report on the evolution of stress activated dynamics in an
aging repulsive colloidal glass. Considerable attention has been
paid to the questions of aging and rejuvenation in these and
analogous systems, as often studied by creep \cite{Cloitre_PRL2000}
or dynamic scattering methods \cite{knaebel2000abl} where
time-elapsed time rescaling is obtained. By contrast, explicit
observations of the loss and storage modes of the system and their
evolution during structural arrest after rejuvenation have not been
made. Specifically, we focus on the regime of stresses below the
viscosity bifurcation where physical aging of the system brings it
to a dynamical arrest on timescales that are comparable to the
observation time. We find that the time evolution of the complex
modulus is strongly dependent on the magnitude of the applied shear
stress. At small stresses, the storage modulus increases and the
loss modulus decreases, both with weak power laws, with $G'>G''$.
The system shows an entirely viscous response in the presence of
large stresses. At intermediate stresses, $G''>G'$ at short times
but develops a peak concomitant with a crossover between $G'$ and
$G''$ as the system ages and undergoes arrest, permitting a novel
observation of the structural relaxation ``in reverse.'' Our results
indicate that this is a characteristic feature of structural arrest
in these systems. The shape of the loss peak is strongly dependent
on the applied stress, with a significant narrowing displayed at
higher stresses.

Our system is an aqueous suspension of Laponite which consists of
disc-like clay particles of 25 nm diameter and 1 nm thickness.
``Wigner'' glasses are formed at low ionic strength in dilute
conditions due to long range electrostatic repulsion between the
particles~\cite{bonn_epl_1998}. The system displays glassy dynamics,
as shown by previous reports
\cite{knaebel2000abl,Bonn_Tanaka_rejuvenation_PRL2002,Joshi_PRE2008}.
Samples are prepared by mixing Laponite XLG (Southern Clay Products)
into ultra-pure water, adjusted to pH 9.5 to ensure chemical
stability of the particles~\cite{mourchid_PRE_1998}. The suspensions
are mixed for 20 minutes and then allowed to develop quiescently for
several days, defining a well controlled and reproducible initial
state. The system is thus in the full aging regime, as opposed to the short time ``gelation'' regime, the delineation between which has been described in the literature\cite{Tanaka_Bonn_PRE2005}. Measurements are conducted on an MCR 301 rheometer
(Anton-Paar) mounted on a air table for mechanical noise isolation.
Evaporation of water from the sample is successfully suppressed
without perturbing the system via application of a thin film of
mineral oil at the sample edge. In all measurements, samples are
first subjected to a rejuvenating shear at $\dot\gamma$=100 s$^{-1}$
for 100 s. This eliminates any flow history and provides a
reproducible initial state in which the viscosity is well defined
within $\pm 5\%$. The sample is then brought to rest and allowed to
sit without external perturbation for 1 second before the start of
subsequent measurements. We monitor the dynamic modulus via
application of a small probe stress oscillation $\sigma_p(t)$ atop a
fixed mean-field stress $\sigma_m$ applied to the sample. The
overall stress is $\sigma(t)=\sigma_m+\sigma_{p_0}\,\sin(\omega t)$
where the applied frequency $\omega$=10 rad/s. Probe stresses were
varied from 0.2 to 1 Pa. In each case, the response of the system
was verified to be linear in $\sigma_{p_0}$, confirming that the
experiment samples the linear properties that exist in the presence
of the mean-field stress $\sigma_m$. Parallel superposition rheology
as applied here has been successfully used to study the dynamics of
a variety of complex fluids under shear. It is to be noted that the
interpretation of results from these flows may be complicated by
coupling between the stationary and oscillatory components and a
shear rate dependence of the perturbation spectrum
\cite{Vermant1998}. This limits strict quantitative interpretation
of data, but not the utility in providing a general description of
the stress or shear rate dependence of dynamics
\cite{DhontWagner_PRE2001}. For this reason, we assess our results
here only qualitatively, as is common practice
\cite{Mewis_Vermant2001,Anderson2006oss}. For observations of the
viscosity bifurcation, following the initialization described, the
sample is subjected to a fixed stress and the time dependent shear
rate developed in response is observed. The sample is re-initialized
and the stress is iteratively increased to cover a suitable range.

\begin{figure}[ht]
\includegraphics[width=80mm, scale=1]{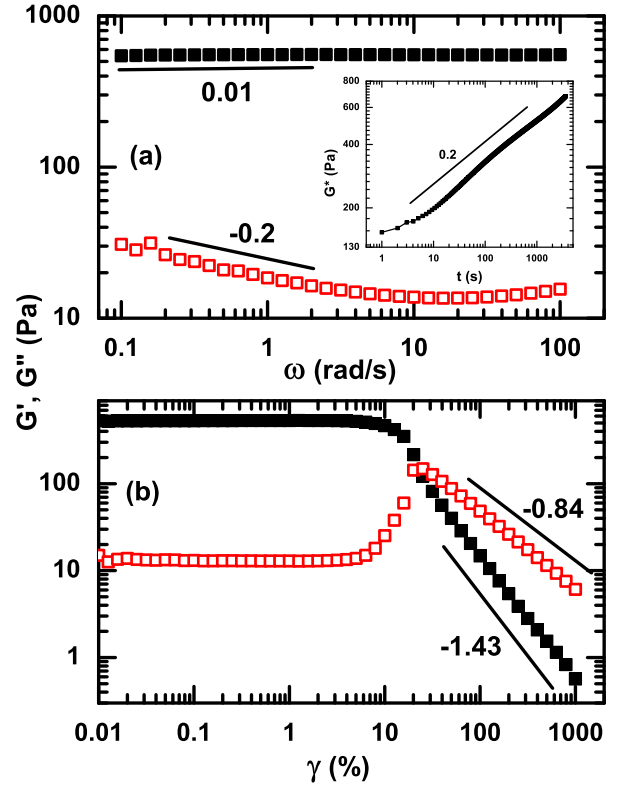}
\caption{Strain and frequency sweeps recorded at $\omega$=10 rad/s.
and $\gamma$=2\%, respectively, for $\varphi$=3.5 wt.\%. Inset: Aging of the complex modulus with time after cessation of the rejuvenating flow, $\omega$=10 rad/s.
\label{sweeps}}
\end{figure}

We examine the frequency dependent and non-linear dynamics of the
system via frequency and strain sweeps. Samples are allowed to sit
quiescently for 30 minutes following the rejuvenating shear at
$\dot\gamma$=100 s$^{-1}$. This duration is large relative to the
sweep time so the system does not age considerably during the actual
measurement. The frequency sweep is typical of this class of
materials, with a wide regime where both the storage and loss moduli
scale weakly with frequency. In this case, $G'\sim\omega^{0.01}$ and
$G''\sim\omega^{-0.2}$, Figure \ref{sweeps}a, for $\varphi$=3.5
wt.\%. In the strain sweep, there is a linear viscoelastic regime of
constant $G'$,$G''$ extending up to $\gamma\approx$ 10\%. This is
followed by a steady decrease in the storage modulus and a peak in
the loss modulus on increasing strain as the system becomes
non-linear, Figure \ref{sweeps}b. Deep into the non-linear regime,
both moduli follow a power-law dependence on $\gamma$ with the ratio
between the exponents approaching 2 with increasing strain, indicative of strong fluidization
of the system. The system displays physical aging with the complex modulus scaling as $G^*\sim t^{0.2}$, after cessation of the rejuvenating shear, inset Figure \ref{sweeps}a.

\begin{figure}[ht]
\includegraphics[width=80mm, scale=1]{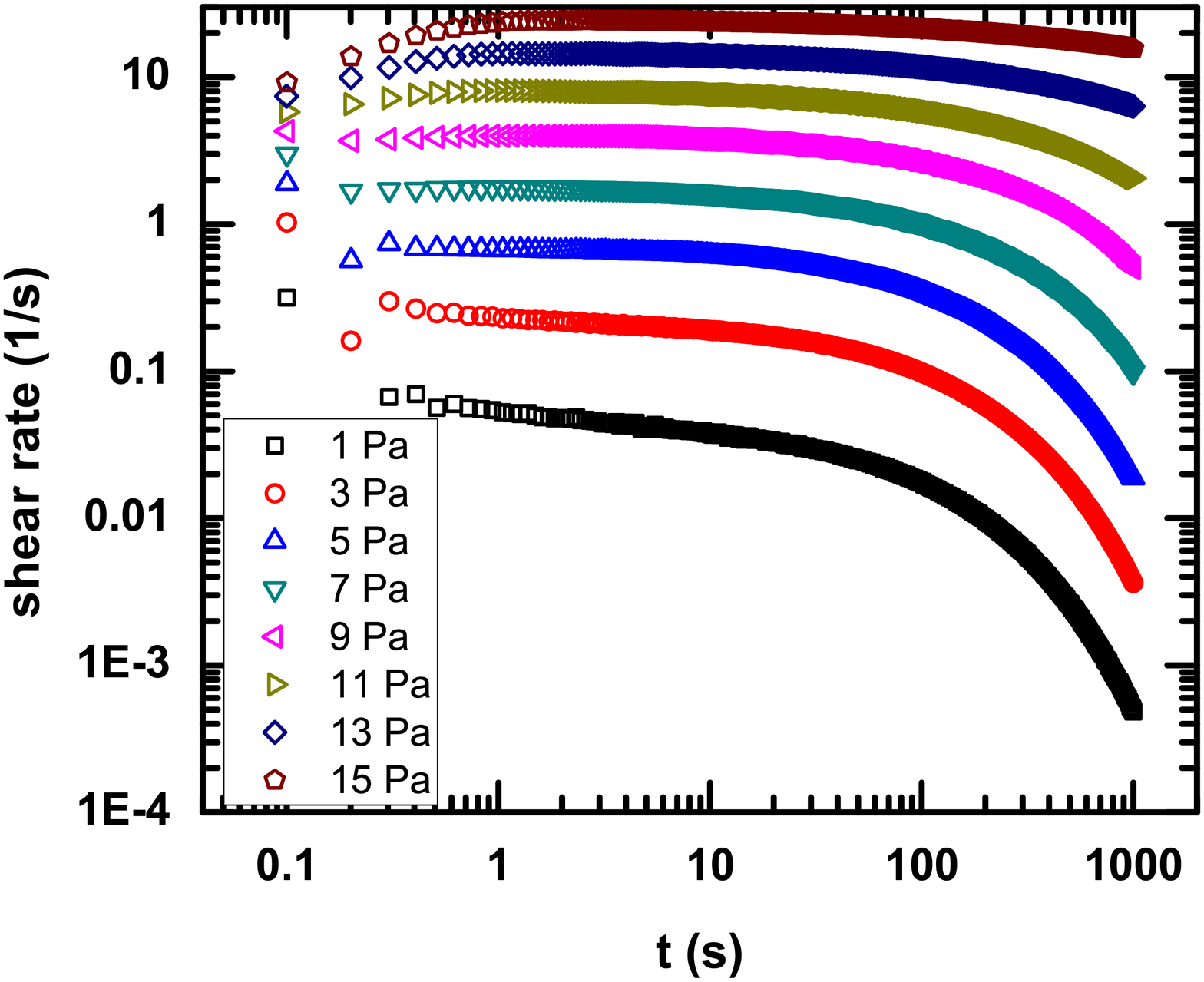}
\includegraphics[width=80mm, scale=1]{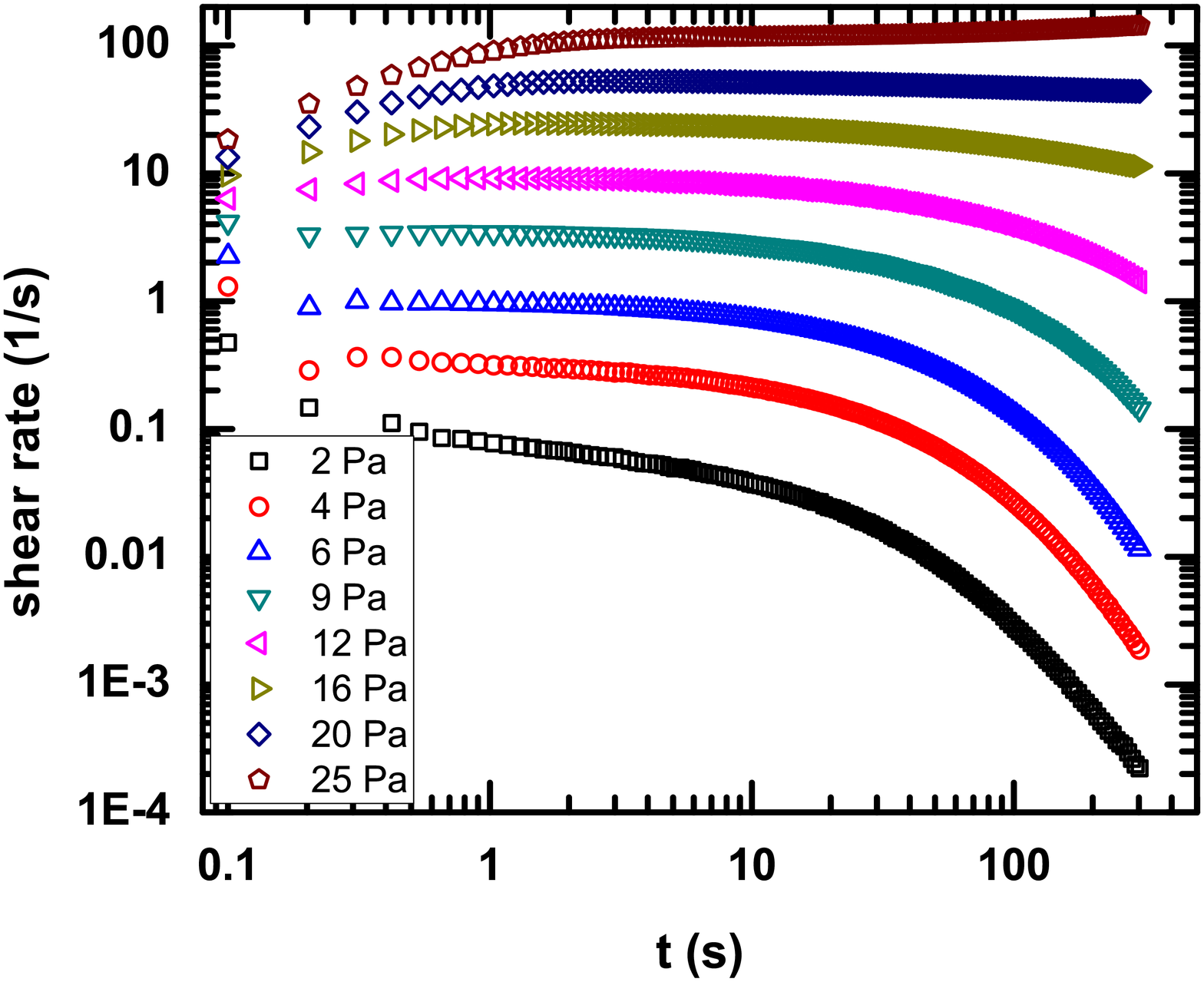}
\includegraphics[width=80mm, scale=1]{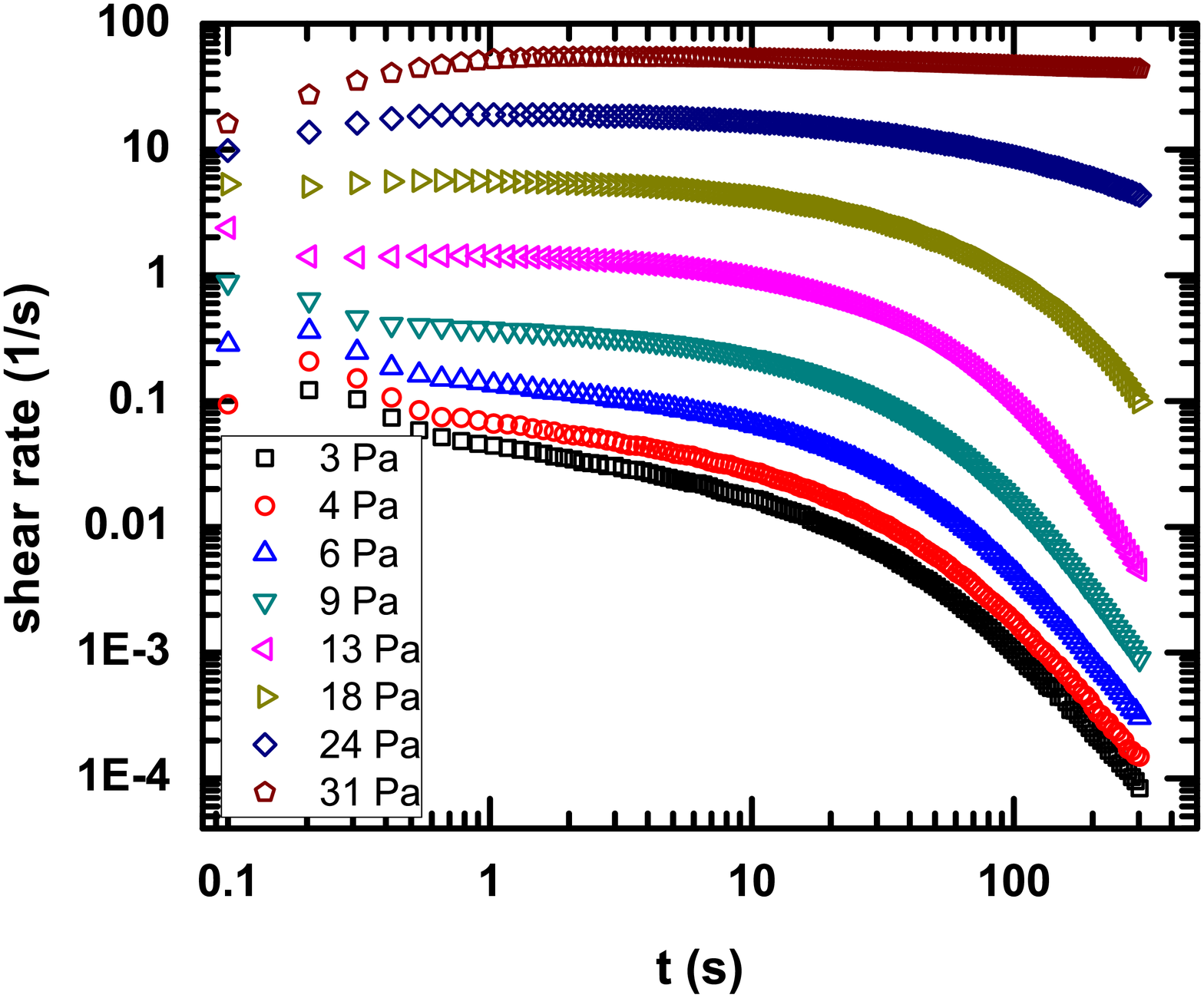}
\caption{Evolution of the shear rate for stresses less than the
bifurcation stress $\sigma_c$ for $\varphi$=2.75, 3.0 and 3.5
wt.\%, (top to bottom).
\label{bifurcation}}
\end{figure}

Application of a fixed stress $\sigma_m$ after sample rejuvenation
produces an initial shear rate which then evolves in time as the
sample ages and its viscosity changes. For stresses larger than some
critical stress, $\sigma_m>\sigma_c$, a steady shear rate was
achieved, indicating that physical aging during the measurement
window was suppressed. Application of smaller stresses
$\sigma_m<\sigma_c$ resulted in a time-dependent decrease of the
shear rate during the course of the measurement. The data show an
initial regime of slow power-law decay followed by a terminal regime
where $\dot\gamma(t)$ appears to diverge towards zero as the
viscosity increases rapidly, Figure \ref{bifurcation}.

The dynamical arrest that occurs in the presence of these
sub-critical mean field stresses is due to the aging of the system
resulting in eventual vitrification. One can think of vitrification
as a structural relaxation conducted ``in reverse.'' In an
equilibrium system such as a model Maxwellian material, the
structural relaxation is traversed by progressively decreasing
frequency. In a non-equilibrium system, observation at a fixed
frequency as a function of time exposes the underlying frequency
dependence as the system ages with the passage of time. The near
linear dependence of the characteristic time on sample age in many
glassy materials which enables time-elapsed time universal scaling
and the width of the glassy power law regime where $G'\sim \omega^x$
with $0<x<1$, both underline this point. One can expect then that
fluidization of a glassy system by an applied stress
$\sigma<\sigma_c$ will reveal a dynamical signature akin to that of
the structural relaxation as the material ages and enters the
arrested state.

\begin{figure}[ht]
\includegraphics[width=80mm, scale=1]{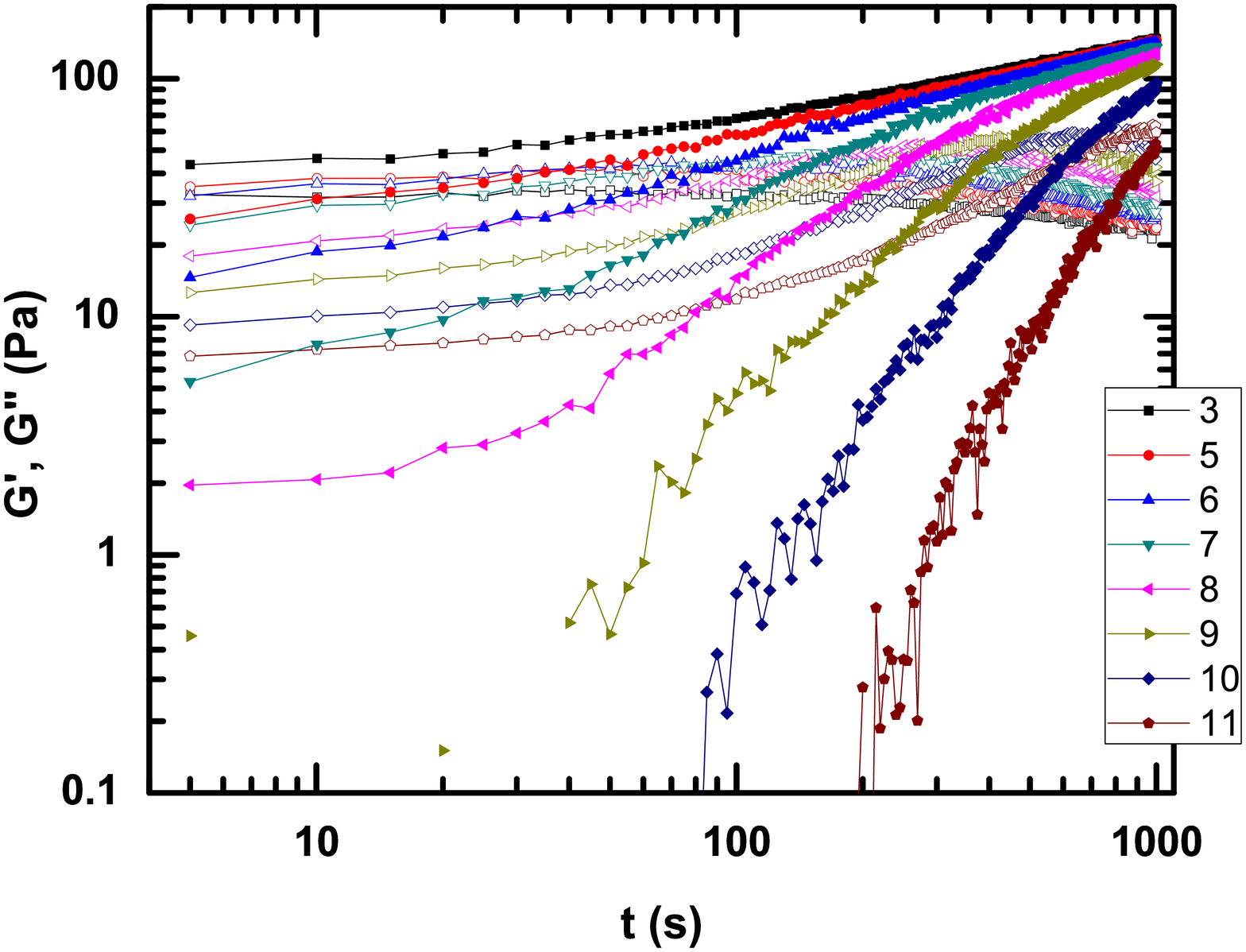}
\includegraphics[width=80mm, scale=1]{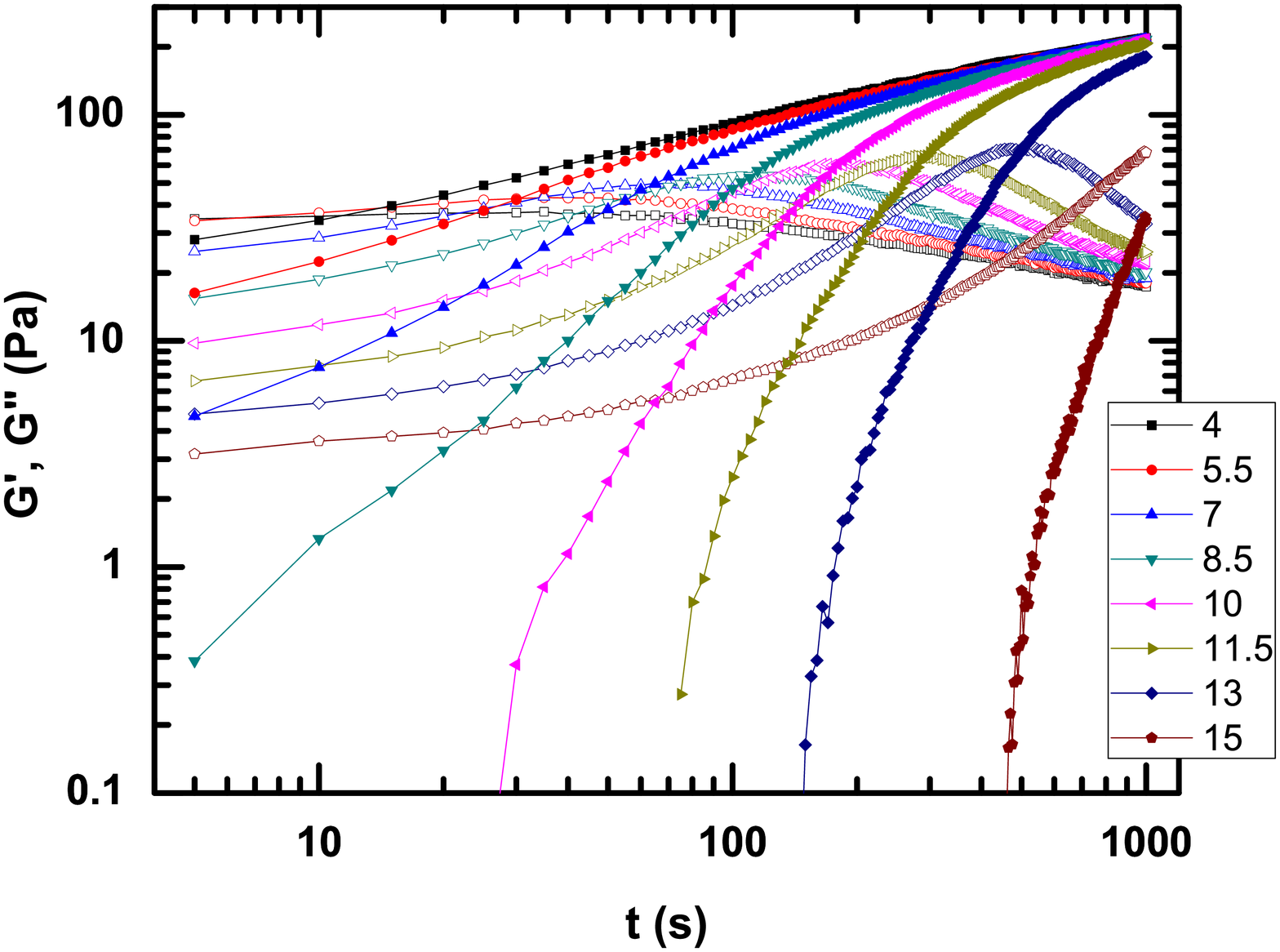}
\includegraphics[width=80mm, scale=1]{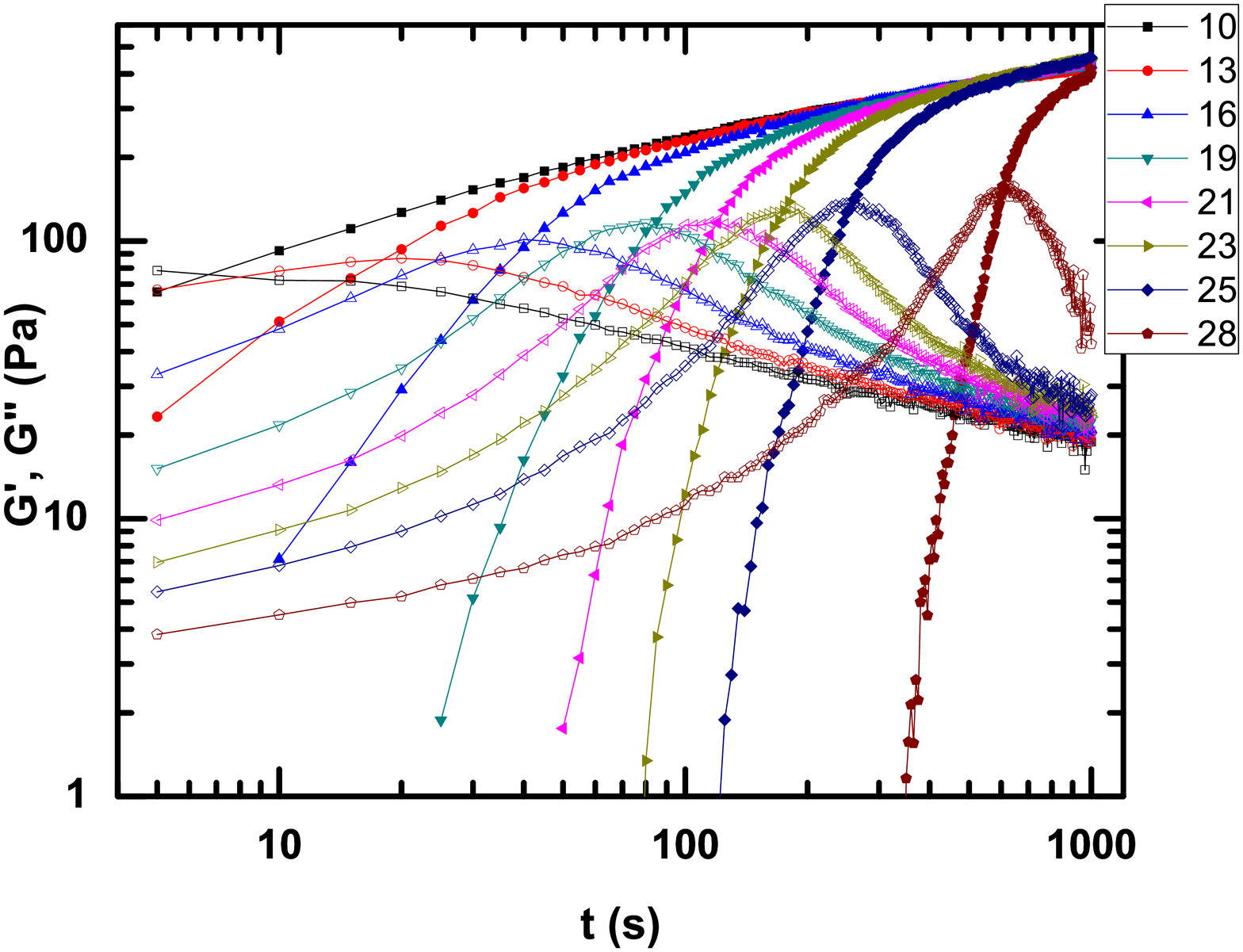}
\caption{Time evolution of storage (solid symbols) and loss moduli
(open symbols), under different applied stresses for $\varphi$=2.75, 3.0 and 3.5
wt.\%, (top to bottom) measured with probe stresses of 0.2, 1 and 1 Pa.
\label{modulus_evolution}}
\end{figure}

\begin{figure}[ht]
\includegraphics[width=80mm, scale=1]{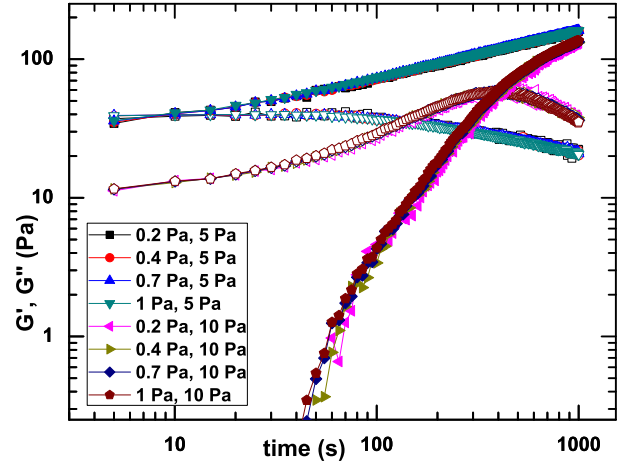}
\caption{Time evolution of storage (solid symbols) and loss moduli
(open symbols), under different probe stresses $\sigma_{p_0}$, for two different mean field stresses (5, 10 Pa) for $\varphi$=2.75 wt.\%.
\label{linearity}}
\end{figure}

This is precisely what is observed. As shown in Figure
\ref{modulus_evolution}, there is a crossover between $G'$ and $G''$
coincident with a peak in $G''$ at a characteristic time
$\tau(\sigma_m)$ during the aging of the system under moderate
mean-field stresses. Similar data have been observed by Ovarlez et al. in their study of a thixotropic bentonite suspension where composition, stress and temperature dependence could be rescaled to generate a master curve of the aging response of the system. \cite{Coussot_PRE2007} The data for the current system are particularly well defined and show striking parallels with the prototypical frequency dependence found in soft systems where there
is a transition via a peaked loss modulus from the terminal regime
where $G'\sim\omega^2, G''\sim\omega$ to a slow arrested state with
weakly increasing $G'$ and decreasing $G''$
\cite{Fielding_Cates_2000}. The linearity of the response to the
probe stress is confirmed by the stress independence of the modulus,
as shown in Figure \ref{linearity}.

\begin{figure}[ht]
\includegraphics[width=80mm, scale=1]{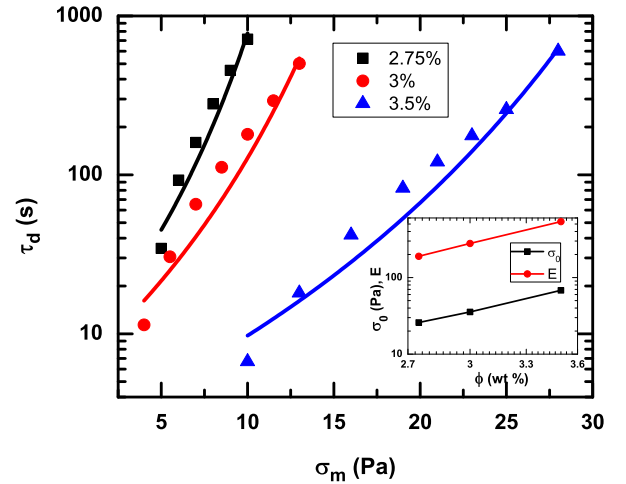}
\caption{Dependence of the crossover time $\tau$ on the mean field
stress $\sigma_m$. The lines are fits to the exponential form
$\tau=\tau_0\,\exp\left[-E/(\sigma_m-\sigma_0)\right]$. Inset: Dependence of fit parameters $E$ and $\sigma_0$ on composition. \label{dynamic_arrest}}
\end{figure}

Notably, $\tau$ appears to vary exponentially with  $1/\sigma_m$.
The data have been fit to Equation \ref{eq:VFT}, as shown in Figure
\ref{dynamic_arrest}, reflecting the slight curvature observed.

\begin{equation}
\tau\sim\exp\left[-E/(\sigma_m-\sigma_0)\right]
\label{eq:VFT}
\end{equation}

In the context of stress as an effective temperature, the timescale
for diffusive particle motion should scale as
$\tau_d\sim\exp(E/\sigma_m)$. Correspondingly, the timescale for
particle arrest should scale inversely, as $\tau_d^{-1}$. The
parameter $\sigma_0$ accounts for the fact that by necessity the
arrest timescale diverges above a critical stress beyond which aging
of the system will not bring it to arrest. This stress should be
identical to the bifurcation stress, $\sigma_c$ discussed earlier.
We observe good correspondence between $\sigma_d$ from the fits of
Figure \ref{dynamic_arrest} and estimates of $\sigma_c$ taken from
the time-dependent shear rate experiments. Both $E$ and $\sigma_0$
exhibit an apparent power law dependence on the volume fraction,
scaling with an exponent $\approx 4$, although the composition range
is quite limited, inset Figure \ref{dynamic_arrest}. More
significantly, however, $E$ scales directly proportional to
$\sigma_0$, indicating that the critical bifurcation stress can be
interpreted in terms of an energetic barrier to particle motion.
This is in agreement with observations which note close
correspondence between bifurcation stresses and the yield stress in
other pasty materials \cite{Coussot_JRheol2002}.

The loss peak that develops as the sample ages and arrests exhibits
a marked dependence on $\sigma_m$, broadening significantly as the
stress is decreased, or conversely, narrowing at higher stresses. A Lorentzian function was used to provide an empirical estimate of the peak width. It was found to decrease markedly from approximately 3 decades for $\sigma_m$=13 Pa to less than 1.4 decades for $\sigma_m$=28 Pa., for $\varphi=3.5$ wt.\%
This narrowing can be understood as due to a narrowing of the
distribution of lossy modes in the glass, as similarly observed in
glassy polymers subjected to external stresses which likewise
activate segmental motion the system \cite{Ediger_Science2009}.
This variation in the shape of the loss peak precludes re-scaling
into a universal form. As a result, the data do not display
time-stress superposition, just as time-temperature superposition is
lost in equilibrium systems on approaching $T_g$ where the width of
the loss modes also become temperature dependent, broadening with
decreasing T \cite{Olsen_PRL2001,oh2002liquid}.

Coussot et al. have advanced a model which describes a time
dependent viscosity resulting from a competition between aging
(structuring) and rejuvenation in the presence of an applied stress
\cite{Coussot_Bonn_PRL2002}. This phenomenological approach
describes the system in terms of a canonical structural parameter
$\lambda$ which grows with a characteristic time, but also is
degraded by the action of shear, with system dependent constants of
proportionality and a specific viscosity that is a simple power-law
function of $\lambda$. Our system displays the hallmark behaviors of
the model such as an asymptotic characteristic time for the onset of
structural arrest at small stresses, an exponential dependence of
this time at intermediate stresses and a fast transient leading to a
steady state viscosity at high applied stresses. Our results here
provide, for the first time, a detailed look at the dynamics of the
system in the low stress regime where arrest occurs and shows that
the features of this data are broadly consistent with observations
made during viscosity bifurcation experiments.

Recent work has shown that the competition between aging and
rejuvenation in soft glassy materials is dependent on the magnitude
of stress or strain imposed \cite{Cloitre_PRL2000}, including the
observation of an over-aging regime \cite{Viasnoff_PRL2002}. Here,
we see that for our repulsive glass, the flow arrest time increases
exponentially, and eventually diverges with increasing stress. We do
not observe an overaging regime within which the flow arrest time
decreases with increasing stress. The reasons for this remain
unclear. One possibility is simply that the stresses used in the
current work are too large for the system under study to elicit an
overaging response. For the range of stresses considered here, the elastic modulus of the colloidal glass was independent of the stress applied during structural arrest. As shown in Figure \ref{modulus_evolution}, within the experimental uncertainty, the system asymptotes towards a single curve, displaying power-law aging of the elastic modulus at long times. This is in contrast to data and a conceptual model derived from bentonite suspensions where both the elastic modulus and the yield strain were seen to increase with the magnitude of the stress applied during the liquid-solid transition\cite{Coussot_PRE2008}. More careful measurements, particularly out to longer times, are required to conclusively test this model in the present system, but it provides a useful starting point for the consideration of the effect of stress on the mechanical properties of the colloidal glass after structural arrest. 

The dynamic measurements demonstrate that the
aging of the system and resulting arrest give rise to a temporal
response that parallels that of the frequency dependence observed in
non-aging systems on traversing the structural relaxation time. The
peak in the time dependent loss modulus is a dynamical signature of
the structural arrest due to aging. Analogous to the manner in which
stress activates particle dynamics in equilibrium systems, it is
shown here that it slows aging, delaying the onset of structural
arrest in this out of equilibrium system. The dependence of the
arrest time on stress is strikingly similar to that of the
relaxation time in thermally activated fragile glass formers on
approaching the glass transition, highlighting the activating role
of stress in structural glasses and further validating the concept
of stress as an effective temperature. To the best of our knowledge,
this represents the first systematic characterization of stress mediated
structural arrest in a colloidal glass due to physical aging. These
results may spur progress on quantitative numerical and experimental
investigation of the role of stress in the aging dynamics of soft
glassy materials.




\acknowledgments
The authors thank P. Coussot and G. Ovarlez for insightful discussions and gratefully acknowledge NSF funding under CBET-0828905.

\bibliography{stress_activated_dynamics}



\end{document}